\documentclass[12pt]{article}

\usepackage{epsfig,multicol,multirow,cite,wasysym}
\usepackage{subfigure,subfigure,latexsym,amssymb}
\usepackage[table]{xcolor}

\newcommand{\be}{\begin{equation}}
\newcommand{\ee}{\end{equation}}
\newcommand{\nn}{\nonumber}
\newcommand{\bea}{\begin{eqnarray}}
\newcommand{\eea}{\end{eqnarray}} 

\newcommand{\vep}{\varepsilon}
\newcommand{\la}{\langle}
\newcommand{\ra}{\rangle}

\newcommand{\Z}{\mathbb{Z}}
\newcommand{\R}{{\kern+.25em\sf{R}\kern-.78em\sf{I} \kern+.78em\kern-.25em}}
\newcommand{\RR}{{\kern+.25em\sf{R}\kern-.6em\sf{I} \kern+.6em\kern-.25em}}
\newcommand{\N}{{\kern+.25em\sf{N}\kern-.78em\sf{I} \kern+.78em\kern-.25em}}
\newcommand{\C}{\mathbb{C}}

\newcommand{\ri}{{\rm i}}

\newcommand{\ltapprox}{\raisebox{-0.5ex}{$\,\stackrel{<}{\scriptstyle\sim}\,$}}

\newcommand{\shi}{\hspace{1.95mm}}

\usepackage{color}
\definecolor{col_red}{rgb}{1.0,0.0,0.0}

\makeatletter
\@addtoreset{equation}{section}
\makeatother

\begin{document}

\begin{flushright}
CERN-TH-2018-189
\end{flushright}
\vspace*{4mm}

\begin{center}
{\Large\bf Topological Susceptibility of the}

\vspace*{6mm}

{\Large\bf 2d O(3) Model under Gradient Flow} \\

\vspace*{1cm}

Wolfgang Bietenholz$^{\rm 1,2}$, Philippe de Forcrand$^{3}$,
Urs Gerber$^{\rm 2}$, \vspace*{2mm} \\
H\'{e}ctor Mej\'{\i}a-D\'{\i}az$^{1}$ and Ilya O.\ Sandoval$^{1}$ \\
\ \\
{\small
$^{1}$  Instituto de Ciencias Nucleares \\
Universidad Nacional Aut\'{o}noma de M\'{e}xico \\
A.P. 70-543, C.P. 04510 Ciudad de M\'{e}xico, Mexico\\
\ \\ \vspace*{-2mm}

$^{2}$ Albert Einstein Center for Fundamental Physics \\
Institute for Theoretical Physics, University of Bern \\
Sidlerstrasse 5, CH-3012 Bern, Switzerland\\
\ \\ \vspace*{-2mm}

$^{3}$ Institut f\"{u}r Theoretische Physik, ETH Z\"{u}rich \\
Wolfgang-Pauli-Str.\ 27, CH--8093 Z\"{u}rich,  Switzerland \\
\ \\ \vspace*{-2mm}
} 
\end{center}

\vspace*{4mm}

\noindent
The 2d O(3) model is widely used as a toy model
for ferromagnetism and for Quantum Chromodynamics.
With the latter it shares --- among other
basic aspects --- the property that the continuum functional
integral splits into topological sectors. Topology can also
be defined in its lattice regularised version, but semi-classical
arguments suggest that the topological susceptibility $\chi_{\rm t}$
does not scale towards a finite continuum limit. Previous
numerical studies confirmed that the quantity $\chi_{\rm t}\, \xi^{2}$
diverges at large correlation length $\xi$.
Here we investigate the question whether or not this
divergence persists when the configurations are smoothened by the
Gradient Flow (GF). The GF destroys part of the topological
windings; on fine lattices this strongly reduces $\chi_{\rm t}$.
However, even when the flow time is so long that the GF impact
range --- or smoothing radius --- attains $\xi/2$, we still 
do not observe evidence of continuum scaling.

\newpage

\tableofcontents

\section{Introduction}

We are going to deal with the 2d O(3) model, a non-linear
$\sigma$-model, which is also known as the Heisenberg model,
or $\C$P(1) model.
It is a highly popular toy model both in solid state physics
--- where it describes ferromagnets --- and in particle physics,
where it shares fundamental features with Quantum Chromodynamics
(QCD). In particular, it is asymptotically free \cite{Poly},
it has a dynamically generated mass gap (which was computed
analytically \cite{massgap} and numerically, {\it e.g.}\
in Ref.\ \cite{Kim}), and in the continuum formulation its
configurations are divided into topological sectors, due
to $\Pi_{2} [S^{2}] = \Z$.

On a square lattice of unit spacing, 
the standard lattice action reads
\be \label{latact}
S [\vec e \, ] = \beta \sum_{\la xy \ra} (1 - \vec e_{x} \cdot
\vec e_{y}) \ , \quad \vec e_{x} \in S^{2} \ , \quad x \in \Z^{2} \ ,
\ee
where $\la xy \ra$ are nearest-neighbour lattice sites.
The na\"{\i}ve continuum limit leads to
\be
S[\vec e \, ] = \frac{\beta}{2} \, \int d^{2}x \, \partial_{\mu} \vec e(x)
\cdot \partial_{\mu} \vec e(x) \ ,
\ee
but at the quantum level it is far from obvious 
if the continuum limit is well-defined.
%how to take a continuum limit and if it is well-defined. 
This is a long-standing issue, which arises in the context of 
topology. In particular, the crucial question is whether or
not the topological susceptibility $\chi_{\rm t}$ exhibits a
continuum scaling behaviour, {\it i.e.}\ whether or not the term
$\chi_{\rm t}\, \xi^{2}$, which is supposed to be a scaling term,
is finite in the continuum limit $\xi \to \infty$
(where $\xi$ is the correlation length in lattice units).

Numerous studies have addressed this question before, including
Refs.\ \cite{BergLuscher,Berg,Luscher82,Pisa92,Michael,Pisa94,Balog}.
After a period of confusion, the consensus seemed to be that 
$\chi_{\rm t}\, \xi^{2}$ diverges at large $\xi$. This conclusion 
is consistent with studies with different lattice actions in the 
same universality class \cite{Blatter,Iperfect,topact}.

The interpretation of this observation is delicate: 
since the functional integral includes all field fluctuations,
the notion of topology is generally non-trivial. For asymptotically 
free theories, one usually refers to the weak-coupling regime,
where one assumes smooth configurations to dominate. For them 
topological sectors are well-defined, along with a % locally defined
topological density $q_{x}$. The topological susceptibility
can be assembled as % from local contributions as
$\chi_{\rm t} = \sum_{x}\la q_{0} q_{x}\ra$, where solely the contact 
term at lattice site $x=0$ causes the divergence \cite{Balog,topact}. 
This implies that it is an UV effect, in agreement with the picture 
of more and more abundant tiny (with respect to $\xi$) topological 
windings as we approach the continuum limit \cite{Luscher82}. 
However, that is not easily compatible with the assumption of 
dominant smooth configurations, and it is questionable if the 
perturbative expansion applies anywhere, even in the UV regime.

We have referred to the usual procedure which invokes the correlation 
length $\xi$ to set a scale. 
It grows exponentially when we increase the inverse coupling $\beta$,
which defines the standard continuum limit, say at fixed physical
size $L/\xi$. However, if this leads to a divergence of 
$\chi_{\rm t} \xi^{2}$, then there is no way to renormalise it
by subtracting counter-terms.
Alternatively, one might consider taking $\chi_{\rm t}$ as a 
reference quantity to fix the scale, and keep $\la Q^{2} \ra = 
L^{2} \chi_{\rm t}$ constant when $\beta$ grows. Then $L$ increases
so slowly that the (standard) size $L/\xi$ shrinks to zero;
this is what it takes to sufficiently damp the
topological fluctuations. Therefore the continuum limit is
ambiguous: there is no way to obtain scaling for 
all observables, which are supposed to be physical. 
This situation is strange and unusual, although 3d U(1) lattice 
gauge theory has an analogous feature \cite{GoMack}, where the 
string tension takes the r\^{o}le of $\chi_{\rm t}$. 

Here we stay with the standard formulation of the continuum limit.
%where (in lattice units) the physical lattice spacing is $1/\xi$.
A conceivable way out is that the divergence of $\chi_{\rm t} \xi^{2}$
is not truly physical, but it can be overcome by systematic
smoothing, which suppresses these tiny topological windings.
% (with a characteristic radius which is small compared to $\xi$). 
Although smoothing techniques have 
been applied before, {\it e.g.}\ in Ref.\ \cite{Michael}, this question 
has not yet been addressed with the systematic
method of the {\em Gradient Flow} (GF). Unlike {\it ad hoc}
approaches, this smoothing procedure is justified based on
the renormalisation group \cite{LuscherGF,LuscherWeisz,Suzuki}.
Here we explore the fate of the term $\chi_{\rm t}\, \xi^{2}$
after applying the GF. 

The GF has a characteristic radius, which we call {\em impact range} 
(cf.\ Subsection 2.4); within this range the configurations become
much smoother, so one might well expect most tiny 
topological windings to be eliminated, and --- possibly ---
the divergence of $\chi_{\rm t}$ as well. 
On the other hand, semi-classical topological windings fulfil the 
equations of motion, so they are not destroyed by the GF. They are 
just a measure-zero subset of all configurations, and the tricky 
question  remains what happens to the frequent tiny winding in the
presence of fluctuations. Hence it is hard to predict whether or not
the divergence of $\chi_{\rm t} \xi^{2}$ will even survive the GF, which 
motivates our numerical study.

There are other models with topological sectors, including
4d SU($N$) gauge theories and QCD,
which suffer from the same problem, if one uses a na\"{\i}ve
lattice formulation of the topological charge density.
However, in QCD this problem is less severe: solutions  without
GF are known \cite{chitQCD}, and the GF provides
another solution \cite{alpha}. In the absence of quarks,
Ref.\ \cite{Ce} discusses in detail how the GF cures
$\chi_{\rm t}$ in 4d SU(3) gauge theory.
This raises the question if this remedy could also cure the
divergence in the 2d O(3) model, which we are going to investigate.

Section 2 describes the numerical tools used in this study.
Section 3 comments on the semi-classical picture, and Section 4
presents our results for the topological scaling behaviour at
the quantum level, based on extensive Monte Carlo simulations.
We discuss the outcome in Section 5. Appendix A compares
various numerical implementations of the GF.
Preliminary results (with short GF times) have been
published in two proceedings contributions \cite{procs}.

\section{Numerical techniques}

\subsection{Algorithm}

Our simulations were performed with the Wolff cluster algorithm
\cite{Wolff89}. It adapts the concept of the Swendsen-Wang algorithm
\cite{SwenWang} from the Ising model to the O($N$) models, where it
is highly efficient. In our study we employed both the
single-cluster as well as the multi-cluster variant.

\subsection{Topological charge}

Regarding the topological charge of a lattice configuration,
we applied the geometric formulation, which
was introduced in Ref.\ \cite{BergLuscher}.
For periodic boundary conditions --- which we assume
throughout this article --- it assigns an
integer charge $Q[\vec e \, ] \in \Z$ to each configuration
(except for a subset of measure zero); this formulation
is reviewed {\it e.g.}\ in Ref.\ \cite{topact},
Section 3.\footnote{Alternative definitions of the topological
  charge of a lattice configuration were suggested in
  Refs.\ \cite{Pisa92,MPV,Iperfect}.}
In the absence of a $\theta$-term, symmetry
implies $\la Q \ra = 0$,\footnote{This can be seen from the
  topological charge density in the continuum,
  $q(x) = \epsilon_{\mu \nu} \vec e(x) \cdot ( \partial_{\mu} \vec e(x)
  \times \partial_{\nu} \vec e(x) )/8\pi \, $: a global
  spin flip $\vec e (x) \to - \vec e (x)$, which is the $Z(2)$
  subgroup of the global O(3) symmetry, changes the sign of $q$.}
hence the topological susceptibility
takes the simple form
\be  \label{topsus}
\chi_{\rm t} = \frac{1}{V} \, \la Q^{2} \ra \ .
\ee

\subsection{Correlation length}

The natural scale of the system is set by the correlation length
$\xi$, which corresponds to the inverse energy gap. It is obtained
by a 2-parameter fit to the correlation function between spin
averages in layers at fixed instances in Euclidean time,
say $x_{2}$ and $x_{2}+r$, with $0 \leq r < L$,
\be  \label{cosheq}
C(r) = \la \vec s_{x_{2}} \cdot \vec s_{x_{2}+r} \ra \propto
{\rm cosh} \Big( \frac{r - L/2}{\xi} \Big) \ , \quad
\vec s_{x_{2}} = \frac{1}{L} \sum_{x_{1}=1}^{L} \vec e_{x} \ ,
\ee
where we refer to a periodic $L \times L$ lattice with sites 
$x=(x_{1},x_{2})$. This proportionality relation holds when
$|r - L/2|$ is sufficiently small. In practice we
follow the recipe of Ref.\ \cite{Wolff90} to determine
$\xi$ by a fit in the range $L/3 \leq r \leq 2L/3$
(varying this range leads to minor modifications of the
fitting result for $\xi$). Our results agree
with the literature, in particular with $\xi$-values
given in Refs.\ \cite{Wolff90,ABF,Kim}.

In each volume $V=L \times L$ we tune $\beta$ such that
$L/\xi \simeq 6$. Hence increasing the lattice
volume corresponds to a system of fixed physical size, which
approaches the continuum limit. The corresponding values of
$\beta$ and $\xi$ are given in Table \ref{tabLbxi}
(the column for $\xi$ at $t=0$ contains the results before
application of the GF).

\begin{table}[h!]
\centering
\begin{tabular}{|r|l|c|c||l|l||l|}
  \hline
$L$~ & ~~~$\beta$ & ${\cal S}_{\xi} ~~ {\cal S}_{\xi_{2}}$ 
& $t_{0}$ & \multicolumn{2}{c||}{$\xi$}
  & ~~~~$\xi_{2}$ \\
  \hline
 & & & & ~~$t=0$ & ~$t=10 \, t_{0}$ & ~~$t=0$ \\
  \hline
  \hline
  24  & 1.263 & 3 ~10 & \shi0.1~~~ & \shi4.03(9) & \shi4.02(5)
              & \shi3.96(1) \\
  36  & 1.370 & 4 ~~5 & \shi0.225 &  \shi5.97(10) & \shi5.96(7)
              & \shi6.01(1) \\
  54  & 1.458 & 5 ~~5 & \shi0.506 & \shi8.95(9) & \shi8.96(7)
              & \shi8.93(4) \\
  80  & 1.535 & 4 ~~5 & \shi1.111 & 12.99(17) & 13.05(11)
              & 13.24(4) \\
  120 & 1.607 & 3 ~~5 & \shi2.5~~~ & 20.14(18) & 19.87(13)
              & 19.77(11) \\
  180 & 1.677 & 3 ~~5 & \shi5.625 & 31.09(36) & 30.39(20)
              & 30.01(18) \\
  270 & 1.743 & 3 ~~5 & 12.656 & 44.80(30) & 45.32(8) 
              & 44.97(24) \\
  404 & 1.807 & 2 ~~5 & 28.336 & 68.34(52) & 67.56(19)
              & 67.66(31) \\
  \hline
\end{tabular}
\caption{Overview of the parameters in our study: we consider
  eight volumes $V = L \times L$, in each one $\beta$ is tuned such
  that $L / \xi \simeq 6$, and the GF time unit amounts to
  $t_{0} = L^{2} / 5760$. When we apply the GF, the correlation
  length $\xi$ does not change significantly up to flow time $10 t_{0}$.
  Before the GF, at $t=0$, $\xi$ agrees fairly well with the
  second moment correlation length $\xi_{2}$, which can be
  measured more precisely. 
  Our numerical measurements are based on ${\cal S}_{\xi}$ and ${\cal S}_{\xi_{2}}$
  independent simulations for $\xi$ and $\xi_{2}$, respectively,
  where each simulation generated $10^{5}$ configurations.}
\label{tabLbxi}
\end{table}

These results are based on sets of ${\cal S}_{\xi}$ independent
measurements, see Table \ref{tabLbxi},
each of which involves $10^{5}$ configurations.
If we insert the errors as obtained from the
fits to eq.\ (\ref{cosheq}), the independent results are
not fully consistent: requiring a unique value at each $L$,
we obtain  $\chi^{2}/{\rm d.o.f.} \simeq 4.0$, which
shows that the errors are underestimated.\footnote{Possible
  reasons are the fixed fitting range, and the use of the same
  configurations to measure the correlation function over all
  distances, although we only include a fixed $x_{2}$ and 
  $r = 0 \dots L-1$ ({\it i.e.}\ one-to-all but no all-to-all 
  correlations).}
Therefore we amplify the errors by a factor of $2$, which leads to
consistency, in particular to $\chi^{2}/{\rm d.o.f.} \simeq 1.0$. 
These extended errors are inserted into the Gaussian propagation 
to obtain the error of the average values given in Table \ref{tabLbxi}.

Despite the sizeable statistics, the uncertainty
in $\xi$ is non-negligible; for comparison, the relative
errors on $\chi_{\rm t}$ are much smaller, see Section 4.
Therefore we also measured the {\em second moment correlation
  length} $\xi_{2}$.
It is obtained from the Fourier transform of the spin-spin
correlation function $\la \vec e_{x} \cdot \vec e_{y} \ra$
at zero momentum ($\chi_{\rm m}$), and at the lowest 
non-zero momentum (${\cal F}$),
\bea
\chi_{\rm m} &=& \frac{1}{V} \sum_{x,y} \la \vec e_{x} \cdot \vec e_{y} \ra
\ , \quad
{\cal F} \, = \, \frac{1}{V} \sum_{x,y} \la \vec e_{x} \cdot \vec e_{y} \ra
\, \cos \Big( \frac{2\pi (x_{1} - y_{1})}{L} \Big) \ , \nn \\
\xi_{2} &=& \frac{1}{2 \sin (\pi /L)} \ \sqrt{\frac{\chi_{\rm m}}{\cal F} -1} \ ,
\qquad (V = L \times L ) \ ,
\eea
where $\chi_{\rm m}$ is the magnetic susceptibility (at magnetisation
zero). $\xi_{2}$ can be measured more precisely than $\xi$, cf.\
Table \ref{tabLbxi}, since it does not require any fit. 
In this case we have performed ${\cal S}_{\xi_{2}}$ independent
measurements, each one based on $10^{5}$ configurations (again
the errors are somewhat enhanced for compatibility of the individual 
results). 
Strictly speaking, this is not the physical scale, but it is known to 
coincide with $\xi$ to high accuracy: in the large-$L$ limit the discrepancy 
is below $1\, \permil$ \cite{CPRV}, and at $L/\xi_{2} \simeq 4$,
$L\geq 70$ it is still below  $1\, \%$ \cite{topact}.\footnote{This
  was observed for the ``constraint lattice action'', cf.\ Section 3.}
(A systematic comparison in other models is given in 
Ref.\ \cite{CaselleNada}.)

\subsection{Gradient flow}

The GF in the O($N$) models has been formulated in Refs.\ \cite{MaSu}.
In the continuum, the spin components $e(x)^{i}$ are altered by
integrating the differential equation
\be  \label{GFcont}
\partial_{t} e(t,x)^{i} = P^{ij}(t,x) \Delta e(t,x)^{j} \ , \quad
P^{ij}(t,x) = \delta^{ij} - e(t,x)^{i} \, e(t,x)^{j} \ ,
\ee
where $\Delta$ is the Laplace operator, and $t$ is the GF time
(of dimension [length]$^{2}$), which starts at $t=0$,
{\it i.e.} $\vec e\,(0,x) = \vec e\,(x)$ and $t \geq 0$.
The GF preserves the spin norm, which corresponds to the condition
$\vec e \cdot \partial_{t} \vec e =0$.

The concept of the GF is based on the heat kernel $K(t,x)$
\cite{LuscherGF,LuscherWeisz,Suzuki},
\be
K(t,x) = \frac{e^{-x^{2}/4t}}{(4 \pi t)^{d/2}} \ ,
\ee
which allows us to estimate its {\em impact range,} 
or smoothing radius, $\bar x(t)$ as (in $d$ dimensions)
\be
\bar x(t) = \left( \int d^{d}x \ x^{2} \, K(t,x) \right)^{1/2}
= \sqrt{2d\, t} \ .
\ee

On the lattice we deal with the spin field $\vec e(t)_{x}$,
and we insert the standard lattice Laplacian,
\be
\Delta \, \vec e(t,x) \ \longrightarrow \
\sum_{\mu =1}^{d} \Big[ \vec e(t)_{x + \hat \mu}
  + \vec e(t)_{x - \hat \mu} \Big] - 2d \, \vec e(t)_{x}
\ , \quad | \hat \mu | = 1 \ .
\ee
For the numerical integration of eq.\ (\ref{GFcont}) also 
the GF time $t$ has to be discretised. Here we apply the 
Runge-Kutta method, see {\it e.g.}\ Ref.\ \cite{NumRec}.
We first compute the gradients to all spin components at all
lattice sites, then we rotate all spins {\em simultaneously} (afterwards
the normalisation $|\vec e_{x}| = 1$ is re-adjusted at each site).
In small and moderate volumes we used the 4-point Runge-Kutta
method, with time step $dt = 10^{-4}$. \footnote{We checked that the
  results coincide within the errors with those obtained at
  $dt = 10^{-5}$. On the other hand, when we increase the step size
  to $dt = 10^{-3}$ we noticed (in a few cases) non-negligible artifacts; 
  they typically emerge at an early stage of the GF.\label{fnRK}}

In this project, the GF integration took most of the computation 
time. In order to handle lattice sizes up to  $L=404$, 
it was mandatory to implement an adaptive step size.
We applied the Dormand-Prince algorithm \cite{DoPri}, which 
gradually increases $dt$, if the Runge-Kutta 4-point and 5-point
gradients agree to high accuracy. At long flow times this method
provides a gain in computing time by several orders of magnitude:
once a configuration is quite smooth, $dt$ can be drastically enhanced
without causing significant artifacts. This is discussed in Appendix A.

In order to compare results in different volumes, and therefore at
different couplings, we need a GF time unit $t_{0}$, which has to be
determined by referring to a dimensional observable. 
Such a time unit allows for the matching of results from different
couplings and volumes, and therefore for a controlled continuum
extrapolation (which is not obvious for {\it ad hoc} smoothing 
techniques). In QCD, $t_{0}$ is usually fixed
by the condition $\la E \ra \, t_{0}^{2} = 0.3$ \cite{LuscherGF}
(or $\la E \ra \, t_{0}^{2} = 0.1$ for SU(2) Yang-Mills theory \cite{HIK}), 
where the density
$E = - {\rm Tr} [G_{\mu \nu} G_{\mu \nu}] /2$
serves as an observable, which is easily measurable ($G_{\mu \nu}$
is a lattice field strength tensor). 

In Refs.\ \cite{procs}
we have used the corresponding density in the 2d O(3) model,
$\la E \ra = \la S \ra / \beta V$.
However, this turned out to be impractical: for increasing GF
time $t$ the (dimensionless) term $\la E \ra \, t$ rises from
0 to some maximum and decreases again. The value of this maximum
decreases as we enlarge $L$, so in order to capture
all volumes under consideration, we had to take a small reference
value like $\la E \ra \, t_{0}^{\rm short} = 0.08$ (for instance,
the value $0.1$ is never attained at $L=404$).
Thus we obtained short time units $t_{0}^{\rm short} \ltapprox 0.1$,
and up to $6 t_{0}^{\rm short}$ the impact range attained
at most $1.6$ lattice spacings.

In order to probe much larger impact ranges, of ${\cal O}(\xi)$,
we now refer directly to $\xi$
as our reference observable to fix $t_{0}$.
We define it such that $10t_{0}$ ---the longest GF time in our
study --- corresponds to an impact range of about $\xi /2$,
\be  \label{t0eq}
t_{0} = \frac{1}{5760} \, L^{2} \simeq \frac{1}{160} \, \xi^{2} 
\quad \longrightarrow \quad
\bar x(10 t_{0}) \simeq \frac{1}{2} \, \xi \ .
\ee
Table \ref{tabLbxi} contains the GF time unit $t_{0}$, as well
as the correlation length measured at $10 t_{0}$; we see that
it hardly changes compared to $t=0$.\,\footnote{In the framework
  of finite temperature gauge theory, Ref.\ \cite{Moore} discusses
  the question how long the GF time can be, before destroying
  physical information. Our results for $\xi (t)$ show that ---
  in our case --- we are on the safe side, at least up to $10 t_{0}$.}

\begin{figure}[h!]
\vspace*{-3mm}
  \begin{center}
\includegraphics[width=0.9\textwidth,angle=0]{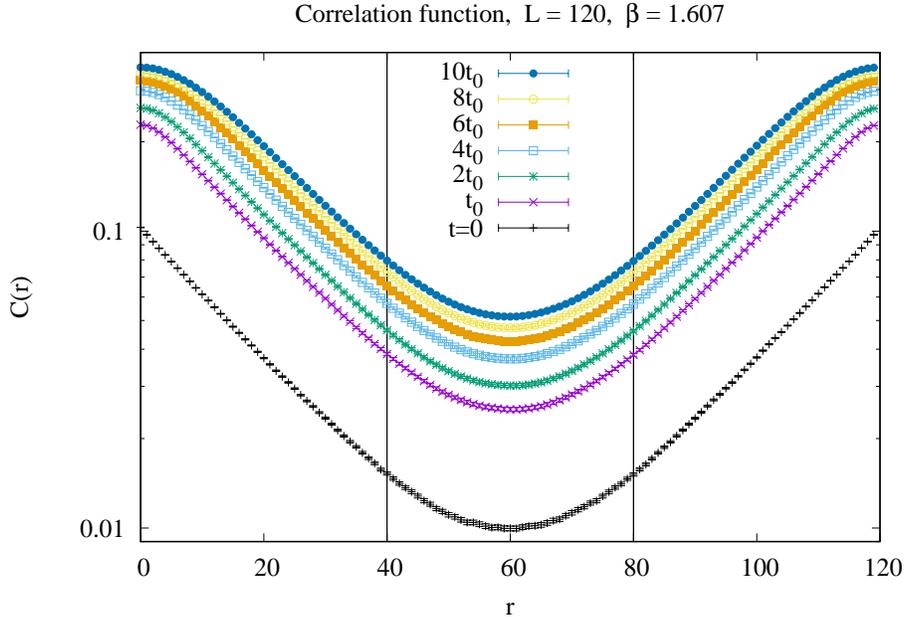}
\end{center}
\vspace*{-6mm}
\caption{The correlation function
  $C(r) = \la \vec s_{x_{2}} \cdot  \vec s_{x_{2} + r} \ra$, measured before
    and after the GF, at $L=120$ (as an example). At fixed separation
    $r$, $C(r)$ rises under GF, but the value of $\xi$ ---
    obtained from a fit to relation (\ref{cosheq}) in the interval
    $40 \leq r \leq 80$ --- remains practically constant.}
\label{correGF}
\end{figure}

An example for the GF time evolution of the correlation function
$C(r)$ of eq.\ (\ref{cosheq}) is illustrated in Figure \ref{correGF}:
at a fixed distance $r$ it increases under GF,
but the value of $\xi$ remains virtually unaffected.
This is consistent with the fact that $\bar x$
is still small compared to $L$, $\bar x(10 t_{0}) \simeq L/12$,
so it does not reach out to the interval, where we performed
fits to relation (\ref{cosheq}).
The GF does, however, have the expected effect of suppressing
the statistical errors in $\xi$ (they are amplified with the
factor of $2$, as at $t=0$, cf.\ Subsection 2.3).

\begin{figure}[h!]
\begin{center}
  \includegraphics[width=0.62\textwidth,angle=270]{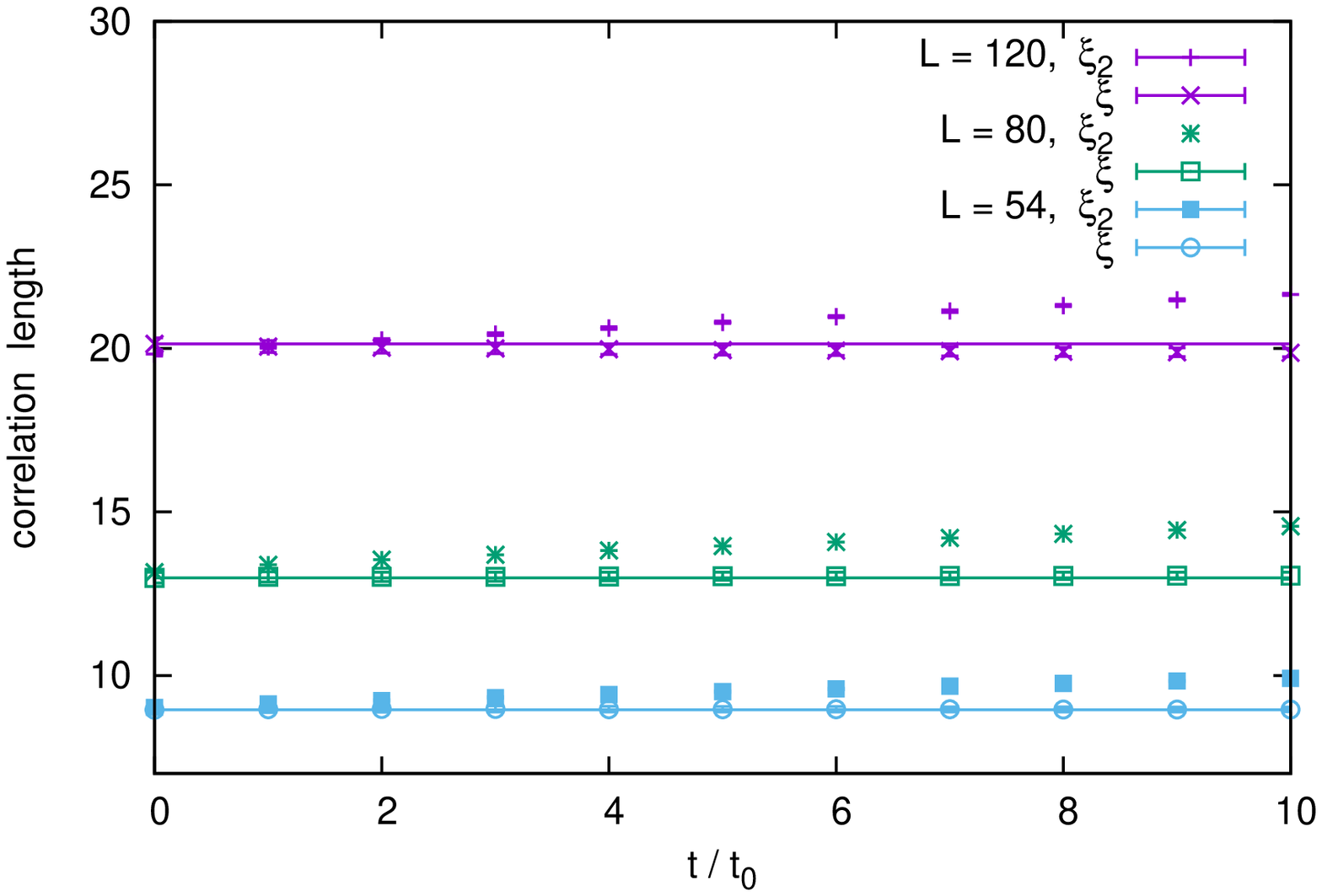}
\end{center}
\vspace*{-6mm}
\caption{A comparison of the correlation length $\xi$, and the
  second moment correlation length $\xi_{2}$, under GF up to
  flow time $10 t_{0}$. We see that they initially agree well, but
  as the GF proceeds, $\xi_{2}$ increases, whereas
  $\xi$ remains stable (this is visualised with horizontal
  lines at the values of $\xi (t=0)$). We show three volumes as
  examples for this generic effect.}
\label{XiXi2fig}
\end{figure}

On the other hand, after applying the GF the second moment 
correlation length $\xi_{2}$ increases above its value at $t=0$, 
as illustrated in Figure \ref{XiXi2fig}.
This property is generic;\footnote{Note that the entire configurations
contribute to the terms $\chi_{\rm m}$ and ${\cal F}$, in contrast 
to the fits, which determine $\xi$ within a limited range. Hence the
short-distance deformation of the correlation function (see Figure
\ref{correGF}) is likely to cause the distortion of $\xi_{2}$.}
it implies that $\xi_{2}(t>0)$ 
cannot be used to set an (approximate) scale. Instead our results
for $\xi (t)$ justify the use of the scale $\xi_{2}(0)$ even after 
the GF, all the way up to $t/t_{0}=10$.

\section{The semi-classical picture}

Ref.\ \cite{BergLuscher} was the first study to show that the
numerical results for the topological susceptibility $\chi_{\rm t}$,
based on Monte Carlo simulations of the standard action (\ref{latact}),
do not seem compatible with continuum scaling, {\it i.e.}\
with the scaling towards a finite continuum limit, which is
na\"{\i}vely expected.
In particular, the dimensionless term $\chi_{\rm t} \, \xi^{2}$
seems to diverge in the continuum limit.
Small topological windings, which may occur in lattice
configurations with low action, were blamed for this effect;
it was suspected that their dominant r\^{o}le on fine lattices
prevents continuum scaling \cite{BergLuscher,Berg}.

Ref.\ \cite{Luscher82} provided a comprehensive semi-classical explanation
for this behaviour. It generally considered 2d $\C{\rm P}(N-1)$
models,\footnote{All the 2d $\C{\rm P}(N-1)$ models, $N=2,3,4 \dots$ , have 
topological sectors (labelled by $Q\in \Z$), in contrast to the 2d O($N$) 
models with $N \neq 3$.} where the continuum instanton action (the minimal 
action within the topological sector $|Q|=1$) amounts to
\be
S_{\rm inst} = \beta \, \epsilon_{\rm inst} \ ,
\quad \epsilon_{\rm inst} = 2 \pi N \ .
\ee
On the lattice, a single topological winding ($Q=\pm 1$) with minimal
action is denoted as a {\em dislocation.} Its action was numerically
obtained as \cite{Luscher82}
\be
S_{\rm disloc} = \beta \epsilon_{\rm disloc} \ ,
\quad \epsilon_{\rm disloc} \simeq 6.69 \cdot N / 2 \ .
\ee
At the quantum level, the fate of a model depends on the balance 
between action and entropy. In this case, a perturbative calculation 
of the $\beta$-function 
suggests that the fate of this model depends on $\epsilon_{\rm disloc}$
as follows \cite{Luscher82},
\be
\epsilon_{\rm disloc} \left\{ \begin{array}{ccc}
  > 4 \pi && {\rm continuum~scaling} \\
  < 4 \pi && {\rm divergence~in~the~continuum~limit.}
  \end{array} \right.
\ee
This implies that continuum scaling of $\chi_{\rm t}$ is
safe at $N \geq 4$. At $N=3$ it can still be arranged for
by adding non-standard terms to the lattice action \cite{Petch}. 
$N=2$, however, is a peculiar case, where
$\epsilon_{\rm inst}$ coincides with the bound derived from
the $\beta$-function. In this case, which corresponds to the
O(3) model, the semi-classical picture predicts the term
$\chi_{\rm t} \, \xi^{2}$ to diverge in the continuum limit.

This semi-classical argument is not rigorous, of course,
there is no compelling reason for it to be conclusive
at the full quantum level. Still, a variety of numerical studies
ultimately suggested that this prediction is confirmed, cf.\ Section 1.

Ref.\ \cite{Blatter} applied a sophisticated lattice action,
a (truncated) {\em classically perfect action,} which was
constructed by means of classical block spin renormalisation group
transformations. It involves couplings over several lattice spacings,
which exclude dislocations with $\epsilon_{\rm disloc} < 4\pi$,
but $\chi_{\rm t} \, \xi^{2}$ still was found to diverge
logarithmically in the continuum limit.

Very different are {\em topological lattice actions,} in particular
the {\em constraint action,} where all configurations have action 0,
if the relative angles between all nearest neighbour spins is
below some bound $\delta$.\,\footnote{If at least one nearest
neighbour angle exceeds $\delta$, then the action is infinite,
{\it i.e.}\ such configurations are excluded from the functional
integral.} In this case, the dislocations are extremely degenerate,
with $\epsilon_{\rm disloc}=0$.
Hence one might expect a very bad divergence of
$\chi_{\rm t} \, \xi^{2}$ in the continuum limit, which is attained
in this case by $\delta \to 0$. It turned out, however, that the
divergence is still compatible with a logarithmic dependence on 
$\xi$ \cite{topact}.

Here this question will be revisited under
application of the GF.\footnote{According to eq.\ (\ref{t0eq})
  we deal with an impact range which is adjusted to $L/12$; 
  it attains $33.7$ lattice spacings in our largest volume. 
  This strongly differs from Refs.\ \cite{procs} (see Subsection 2.4),
  and also from Ref.\ \cite{Blatter}, where the coupling range of
  the ``perfect lattice action'' was fixed to a couple of lattice
  spacings, while $\xi$ increased up to $58$.}
Before doing so, however, we begin with an observation about
the relevance of the semi-classical picture.
To this end, it is sufficient to consider modest lattice volumes, 
of sizes $L=24 \dots 80$, with the $\beta$-values of
Table \ref{tabLbxi}.
In each volume we selected 50\,000 configurations
with topological charge $|Q| = 1$.

\begin{figure}[h!]
\includegraphics[width=0.6\textwidth,angle=270]{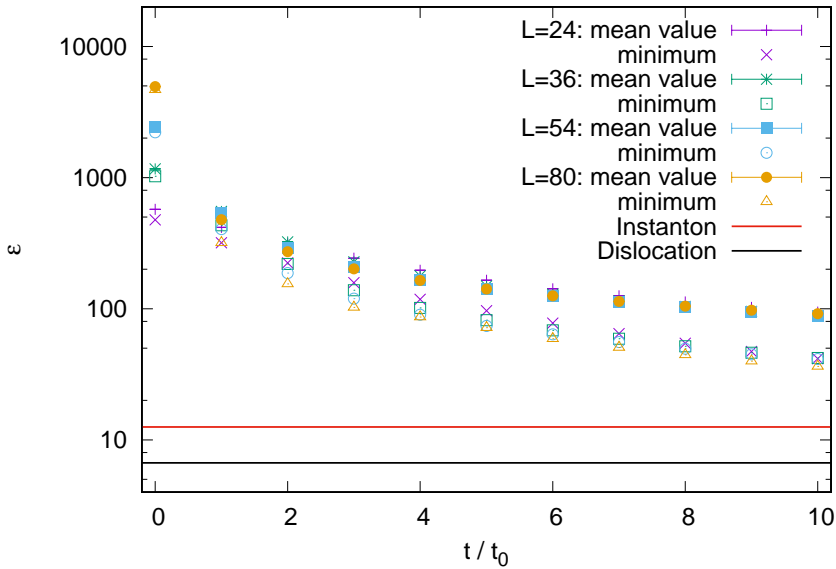}
\vspace*{-3mm}
\caption{The quantity $\la \epsilon \ra = \la S \ra /\beta$
measured in volumes $V= L\times L$, at $L/\xi \approx 6$.
In each volume, and at each flow time $t/t_{0} = 0,1 \dots 10$,
we used 50\,000 configurations with topological charge $|Q|=1$.
We also show the minima of $\epsilon$ within this set.
Even after a long GF, up to flow time $10 t_{0}$, these minima
are still more than a factor of $5$ above the dislocation
value $\epsilon_{\rm disloc} \simeq 6.69$. We infer that dislocations 
and their vicinities hardly contribute to the statistics.}
\label{semifig}
\end{figure}

Figure \ref{semifig} refers to the quantity $\epsilon = S /\beta$
($S$ being the lattice action (\ref{latact})): it shows the mean
value $\la \epsilon \ra$, as well as the minimum obtained in each
volume. At GF time $t=0$ even
the minima (in this set of configurations) 
are orders of magnitude above the instanton and
dislocation values. This suggests that --- although configurations
with $\epsilon$ down to $\epsilon_{\rm disloc}$ exist --- their
contribution to a typical expectation value is negligible
in our settings.\footnote{Actually such configurations
have the highest probability 
$p[\vec e \, ] \propto \exp(-S[\vec e \, ])$,
but configurations with a
significantly larger action have a much higher degeneracy, such
that they overwhelmingly dominate the functional integral.}

When we apply the GF, as described in Section 2, the configurations
become smoother and the action decreases, so one might suspect
that now the semi-classical configurations (or at least their
vicinity) become relevant. Figure \ref{semifig} shows
that this is {\em not} the case: even when we run the GF
up to $10 t_{0}$, the averages and minima
(within a set of 50\,000 configurations, at any instant $t$) are
still more than a factor of $5$ times larger than $\epsilon_{\rm disloc}$.

Figure \ref{semifig} further shows that this observation hardly
depends on the volume. It raises the question how relevant
the semi-classical consideration really is, since it does
not refer to the statistically significant contributions 
(unless presumably in tiny physical volumes).
Nevertheless, our goal is a direct verification of its prediction; 
this is the question to be addressed in the next section.

\section{Topology under the Gradient Flow}

Based on ${\cal S}_{\chi}$ sets of $10^{5}$ configurations in each
volume (see Table \ref{tabchit}), we finally measured the topological
susceptibility $\chi_{\rm t}$, given in eq.\ (\ref{topsus}).
Unlike the case of $\xi$, the results for $\chi_{\rm t}$
from these ${\cal S}_{\chi}$ independent simulations are consistent 
within our estimated errors: in this case we obtain a ratio
$\chi^{2}/{\rm d.o.f.} \simeq 0.90$ (the cluster algorithm
allows us to avoid topological auto-correlations). 

\begin{figure}[h!]
\begin{center}
\includegraphics[width=0.6\textwidth,angle=270]{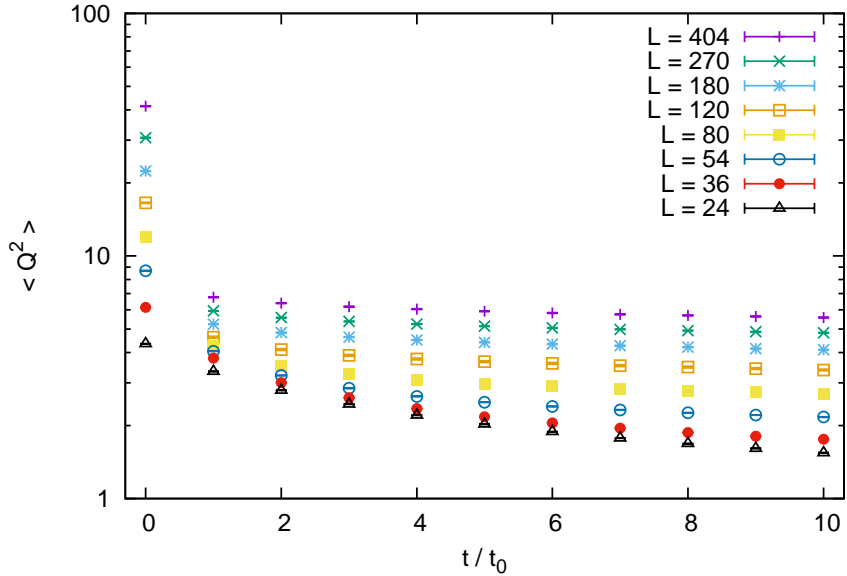}
\end{center}
\vspace*{-6mm}
\caption{GF time evolution of the expectation value
$\la Q^{2} \ra = \chi_{\rm t} V$.}
\label{Q2vsGF}
\end{figure}

Our results for $\chi_{\rm t}$ are listed in Table \ref{tabchit}.
They are averaged over all simulations, 
and each of their standard errors 
enters the Gaussian composition of the final error.
The evolution under GF is illustrated in Figure \ref{Q2vsGF}, which
shows the dimensionless term $\la Q^{2} \ra = \chi_{\rm t} V$.
In large lattice volumes, {\it i.e.}\ on fine lattices, 
we see a rapid decrease of $\la Q^{2} \ra$ when the GF starts, 
in particular from $t=0$ to $t_{0}$ (in Appendix A we will see 
that most of this effect happens even within a first small 
fraction of $t_{0}$). At a later stage  $\la Q^{2} \ra$ still
keeps decreasing, but at an ever slower rate.

\begin{table}[h!]
\centering
\begin{tabular}{|r||l|l|l|l|l|}
  \hline
$L$~~ ${\cal S}_{\chi}$& \multicolumn{5}{c|}{$\chi_{\rm t}$ \ (in units of $10^{-3})$} \\
  \hline
  & ~~$t=0$ & ~~~$t_{0}$ & ~~$2 \, t_{0}$ & ~~$5 \, t_{0}$ & ~~$10 \, t_{0}$ \\
  \hline
  \hline
 24 ~~5 & 7.54(1)   & 5.80(1)   & 4.85(1)   & 3.516(7)  & 2.681(6)  \\
 36 ~~5 & 4.736(9)  & 2.926(6)  & 2.319(5)  & 1.677(3)  & 1.356(3)  \\
 54 ~~5 & 2.982(7)  & 1.388(3)  & 1.103(2)  & 0.856(2)  & 0.743(1)  \\
 80 ~~5 & 1.87(1)   & 0.662(2)  & 0.552(1)  & 0.466(1)  & 0.423(1)  \\
120 ~~5 & 1.150(6)  & 0.321(3)  & 0.287(2)  & 0.255(2)  & 0.235(2)  \\
180 ~~3 & 0.691(2)  & 0.1614(4) & 0.1491(4) & 0.1360(4) & 0.1266(4) \\
270 ~~3 & 0.422(1)  & 0.0843(2) & 0.0765(2) & 0.0705(2) & 0.0662(2) \\
404 ~~2 & 0.2538(8) & 0.0414(1) & 0.0392(1) & 0.0362(1) & 0.0342(1) \\
  \hline
\end{tabular}
\caption{The topological susceptibility $\chi_{\rm t}$ of
eq.\ (\ref{topsus}) on $L \times L$ lattices, with the values
of $\beta$ and $t_{0}$ given in Table \ref{tabLbxi}, based on
${\cal S}_{\chi}$ measurements with $10^{5}$ configurations each.}
\label{tabchit}
\end{table}

At last we arrive at the discussion of the ``scaling term''
$\chi_{\rm t}\, \xi^{2}$.
Regarding the correlation length, we rely on the property that
the flow times are not excessively long, so that physical
aspects are not affected, and in particular the long-range
scale $\xi$ should not change, as we argued in Subsection
2.4. In fact, our results in Table \ref{tabLbxi} confirm that 
the modifications of $\xi$ are minor: in each volume, 
$\xi (0)$ and $\xi (10 t_{0})$ agree within less than 
$1.7 \sigma$. (It is also noteworthy that the sign of 
$\xi (0) - \xi(10 t_{0})$ differs in our results from
different lattice volumes, which further shows the absence
of a systematic effect of the GF on $\xi$ up to $10 t_{0}$.)

\begin{figure}[h!]
\begin{center}
\vspace*{-5mm}
\includegraphics[width=0.73\textwidth,angle=270]{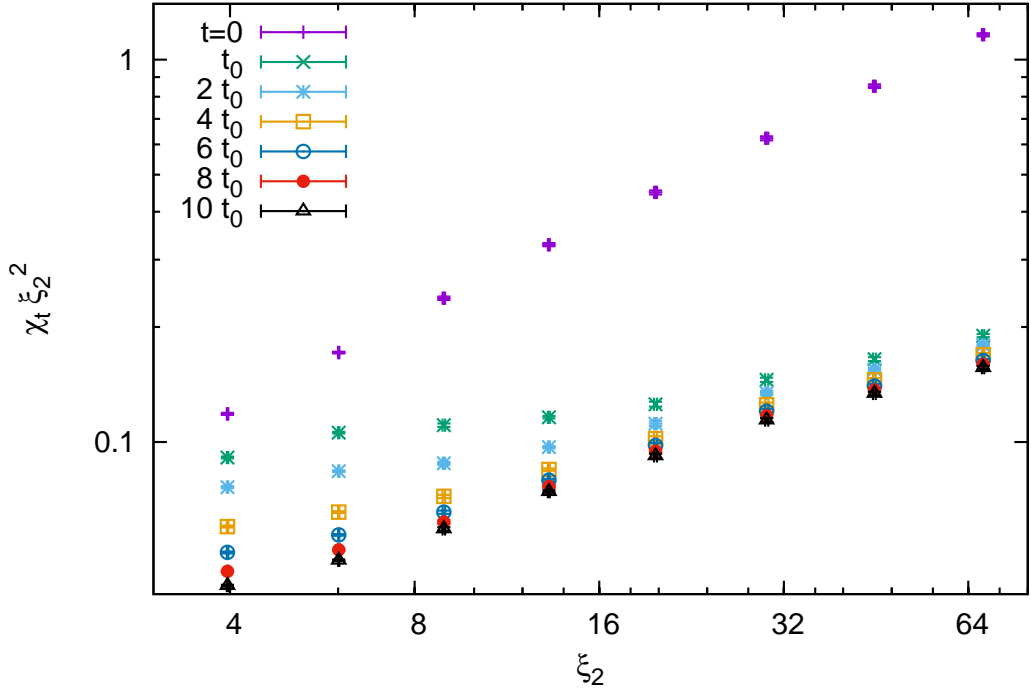}
\end{center}
\vspace*{-7mm}
\caption{Illustration of the non-scaling of the term $\chi_{\rm t}\,\xi^{2}$.
The scale used in this plot is the second moment correlation length $\xi_{2}$.}
\label{scaleGF}
\end{figure}

Trusting the stability
of $\xi$, we replace it by $\xi_{2}(0)$, for which we have
precise results --- see Table \ref{tabLbxi} --- and use it at
any flow time $t \in [0,\, 10 t_{0}]$, cf.\ Subsection 2.3.
This yields the scaling plot in Figure \ref{scaleGF}.

It is an unambiguous observation that --- after any fixed multiple 
of the flow time unit $t_{0}$ that we considered --- the quantity
$\chi_{\rm t}\, \xi^{2}$ keeps growing as we increase the
correlation length;
we cannot observe convergence towards a finite continuum limit.
This trend is most obvious in our largest lattice volumes and 
at long flow times. At relatively
short GF, in particular at flow time $t_{0}$, the ``scaling term''
looks almost stable up to $L=80$, $\xi \approx 13$, but even
closer to the continuum limit it turns into the (qualitative)
behaviour observed at long flow times.

As a first hypothesis, we assume the asymptotic behaviour at large 
$\xi$ to be logarithmic.\footnote{Here and in the following we refer
to $\xi$, which is conceptually correct, although in practice it is 
replaced by $\xi_{2}$, as we explained before.} 
This can be expressed by the ansatz
\be  \label{logfun}
\chi_{\rm t}\, \xi^{2} = a_{1} \ln (a_{2} \, \xi + a_{3}) \ ,
\quad a_{i} = {\rm constant} \ ,
\ee
which was successful in fits to results obtained with topological 
lattice actions \cite{topact}. As an alternative, we consider 
another 3-parameter ansatz, which describes a power-law,
\be  \label{powfun}
\chi_{\rm t}\, \xi^{2} = b_{1} \xi^{b_{2}} + b_{3} \ , \quad
b_{i} = {\rm constant} \ ,
\ee
as in Refs.\ \cite{topact,procs}. That behaviour corresponds to the
semi-classical picture of Ref.\ \cite{Luscher82} (for the case of
4d Yang-Mills gauge theory, this property is worked out explicitly
in Ref.\ \cite{UJW90}).

We first consider the data before the GF. In this case,
we perform fits over the entire range $L= 24 \dots 404$, 
so there are 5 ``degrees of freedom'', and we obtain at $t=0$
\bea
a_{1} = 1.3(1), \ a_{2} = 0.021(3),
\ a_{3} = 1.014(2), && \chi^{2}/{\rm d.o.f.} = 3.76 \ \nn \\
b_{1} = 0.0522(5), \ b_{2} = 0.741(2),
\ b_{3} = -0.026(1), && \chi^{2}/{\rm d.o.f.} = 0.04 \ . \nn
\eea
\begin{figure}[h!]
\begin{center}
\hspace*{-5mm}
\includegraphics[width=0.54\textwidth,angle=270]{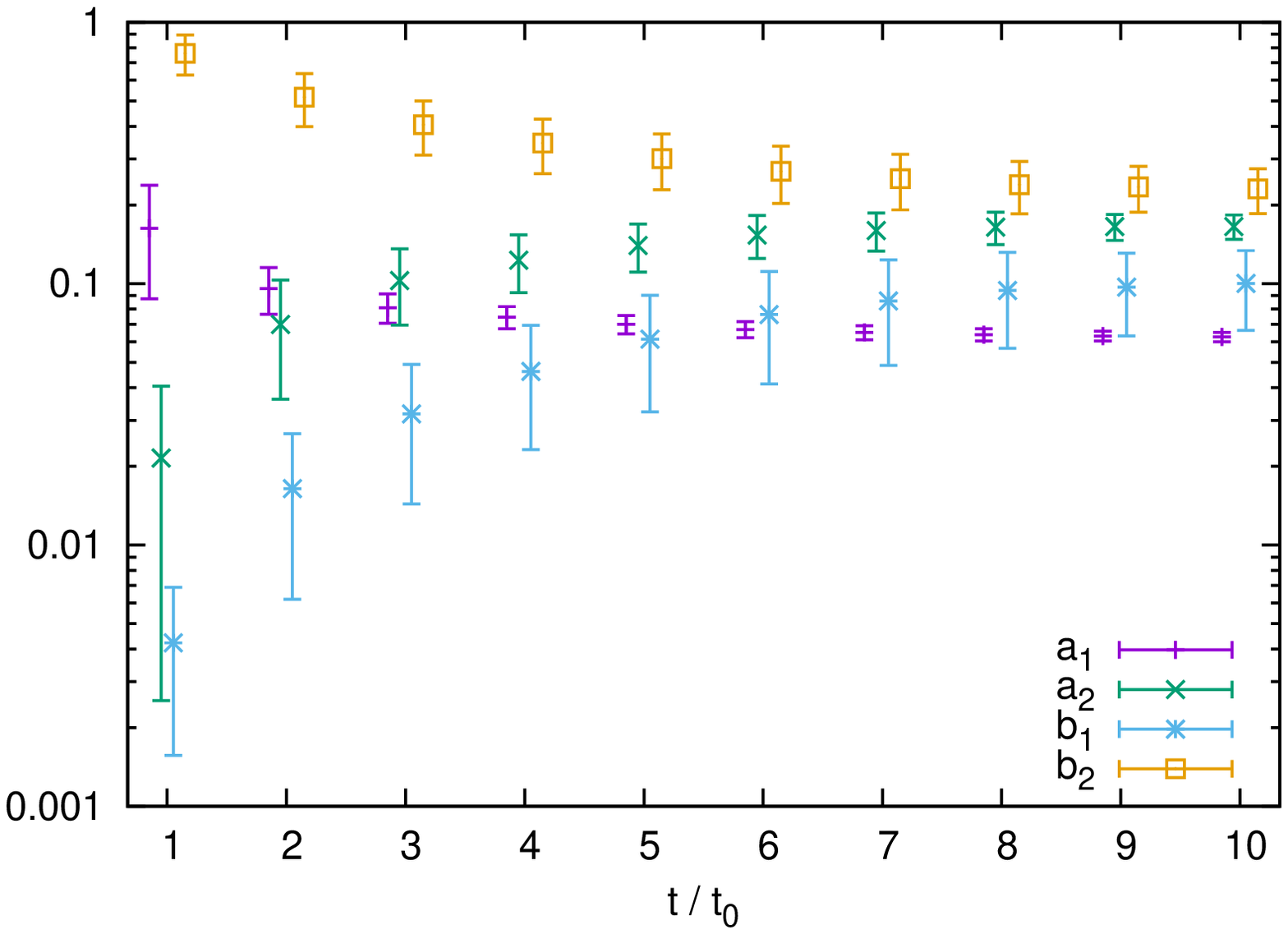} \\
\includegraphics[width=0.52\textwidth,angle=270]{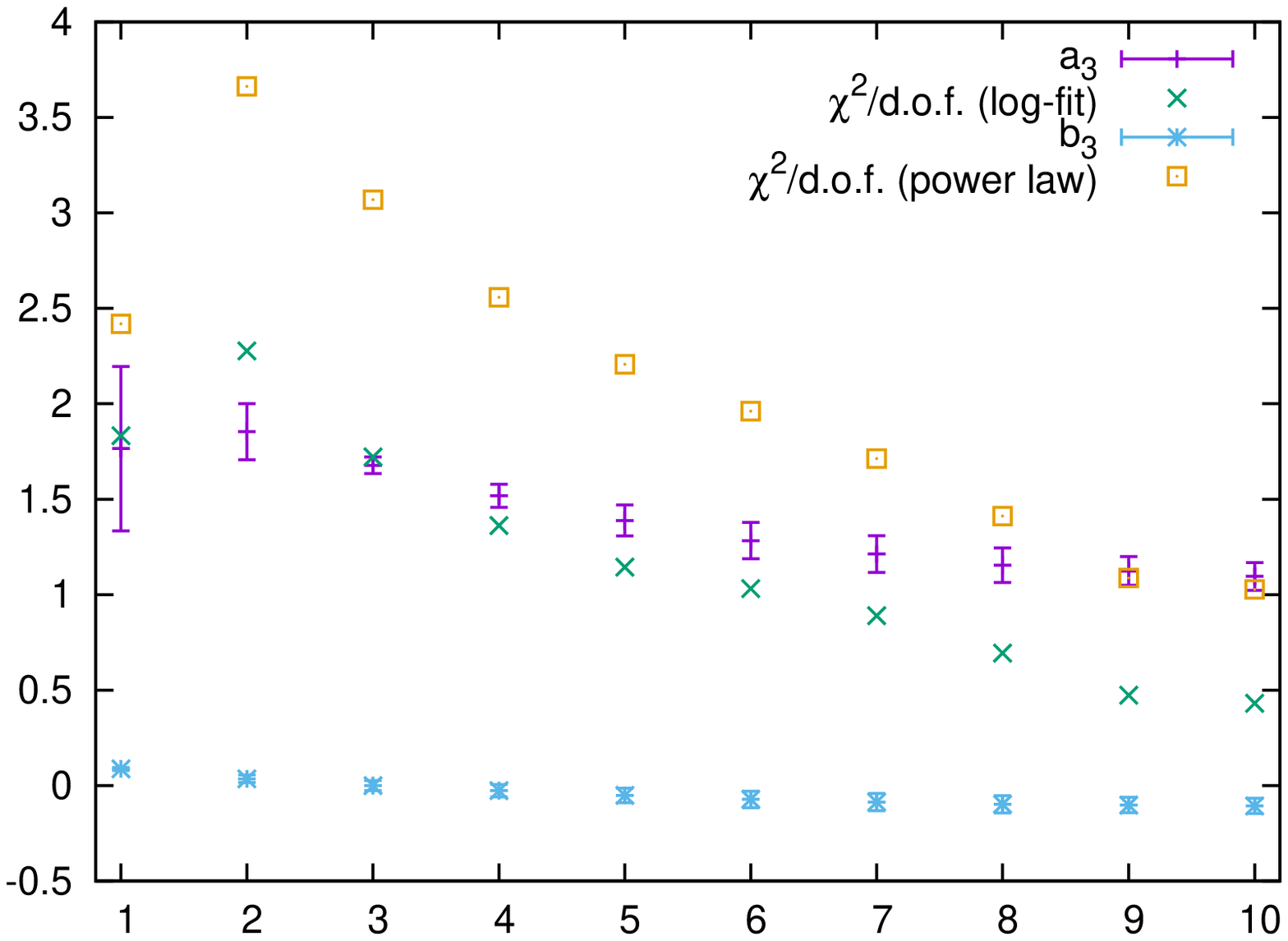}
\end{center}
\vspace*{-3mm}
\caption{The constants $a_{i}(t)$ and $b_{i}(t)$ of the fits to the
logarithmic function (\ref{logfun}) and to the power-law (\ref{powfun}),
respectively, along with the $\chi^{2}/{\rm d.o.f.}$-ratios. 
The fits were performed at any flow time $t/t_{0} = 1 \dots 10$, 
in the range $L=54 \dots 404$ (${\rm d.o.f.}=3$).
In all cases the logarithmic fits have a better quality, in contrast
to the case $t=0$. The constants $a_{1},\, a_{2},\, b_{1},\, b_{2}$ are 
positive and incompatible with 0 at any GF time up to $10 t_{0}$.
Hence our data are incompatible with continuum scaling.} 
\label{constGF}
\end{figure}
The power-law fit has a tiny value of $\chi^{2}/{\rm d.o.f.}$
(which appears accidental), but this quantity is somewhat 
large for the logarithmic fit.\footnote{The fits to our 
preliminary data that we considered
in Refs.\ \cite{procs} had a similar quality for both functions.
On the other hand, in Ref.\ \cite{topact} we observed superiority
of the logarithmic function for data obtained with the 
constraint lattice action.}
However, even there the uncertainties of the fitting parameters 
are moderate. 
The observation that the constants 
$a_{1}, \, a_{2}, \, b_{1}, \, b_{2}$ are all
larger than 0 (far beyond the errors) confirms that the 
data before GF are incompatible with continuum scaling.

Figure \ref{constGF} shows the constants $a_{i}(t)$ and $b_{i}(t)$ obtained
from the fits to the functions (\ref{logfun}) and (\ref{powfun}) at
GF times $t = t_{0}, \, 2t_{0} \dots 10t_{0}$. The lower plot also
shows the ratio $\chi^{2}/{\rm d.o.f.}$, as a measure of the
quality of the fits. All fits were performed in the range
$L=54 \dots 404$, hence they capture 6 data points, corresponding 
to 6 lattice volumes. 
In all these cases, {\it i.e.}\ after the GF, the fits to the logarithmic
ansatz (\ref{logfun}) are superior, as we see from the lower plot in 
Figure \ref{constGF} (this behaviour agrees with Refs.\ \cite{procs}).

The essential observation, however, is based on the upper plot:
it shows that the constants $a_{1},\, a_{2},\, b_{1},\, b_{2}$ 
keep on being larger than zero during the GF;
zero values are well beyond the errors.\footnote{In light of 
Figure \ref{constGF}, one might question the behaviour of $a_{2}$ 
and $b_{1}$ at short $t \leq t_{0}$. However, in Refs.\ \cite{procs} 
we arrived at the same conclusion also for such short GF times.}
Therefore, even after the GF our data are incompatible with a 
scaling of $\chi_{\rm t} \xi^{2}$ towards a finite continuum limit.

\section{Conclusions}

There is a variety of models with topological sectors, and
some of them are plagued by problems with the continuum scaling
of $\chi_{\rm t}$. This is not the case in the simple 1d O(2)
model, where --- for a multitude of lattice actions ---
$\chi_{\rm t}\, \xi$ exhibits a straight convergence to its
continuum value of $1/2\pi^{2}$ \cite{rot97,topact}.

In na\"{i}ve lattice formulations of 4d Yang-Mills gauge theory,
as well as QCD, this problem appears, but there are various
ways to overcome it, see Ref.\ \cite{LuscherPalombi} for pure
SU(3) gauge theory, and the aforementioned Refs.\
\cite{chitQCD,alpha} for QCD.

Regarding the 2d $\C$P($N-1$) models, the numerical results
confirm the semi-classical picture of Ref.\ \cite{Luscher82}
that we sketched in Section 3: no problem occurs at $N \geq 4$,
and at $N=3$ there is a divergence, but it can be avoided by
non-standard lattice actions, see {\it e.g.}\ 
Refs.\ \cite{Petch,Ruedi}.

There remains the case $N=2$, which is peculiar indeed:
in this model, which is equivalent to the 2d O(3) model,
no way around the divergent continuum limit of $\chi_{\rm t}\, \xi^{2}$
is known; we have seen that not even the GF, which is a safe
remedy in other models, helps in this specific case.

\begin{figure}[h!]
\vspace*{-9mm}
\begin{center}
\includegraphics[width=0.43\textwidth,angle=270]{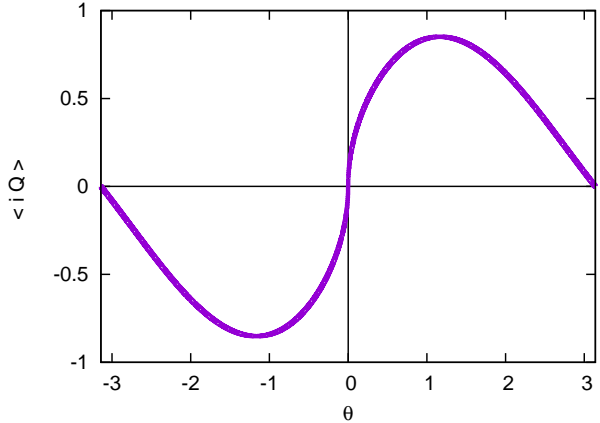}
\end{center}
\vspace*{-3mm}
\caption{A schematic illustration of the expectation value
  ${\rm i} \la Q \ra$ as a function of the vacuum angle 
  $\theta$, in the continuum limit.
  The peculiarity of the 2d O(3) model is that its slope
  --- which is proportional to $\chi_{\rm t}$ --- seems to diverge
  at $\theta =0$. This picture corresponds to Ref.\ \cite{Bogli},
  and to private communication with A.\ Zamolodchikov;
  our results suggest that its qualitative features persist
  under the GF.}
\label{Qtheta}
\end{figure}

This does not mean that {\em all} topological terms
in the 2d O(3) model are ill-defined. Even without GF, 
there is evidence for the opposite to hold for the 
following quantities:

\begin{itemize}

\item The correlation function of the topological charge density
  $q_{x}$, $\la q_{x} q_{y}\ra$, is well-defined ({\it i.e.}\ finite
  in the continuum limit) at all separations $x-y$, expect
  for $x=y$. That point alone causes the divergence of
  $\chi_{\rm t} = \sum_{y} \la q_{x} q_{y}\ra$ \cite{Balog,topact},
  and the situation is similar in QCD \cite{chitQCD}.\footnote{In a
  fixed topological sector, the correlation $\la q_{x} q_{y}\ra$ at 
  large separation can be employed for an indirect measurement 
  of $\chi_{\rm t}$ \cite{AFHO}.}
 
\item The kurtosis $c_{4} = (3 \la Q^{2} \ra^{2} - \la Q^{4} \ra)/V$
is a characteristic of the distribution of the topological charges
(it vanishes if this distribution is Gaussian). In the continuum limit, 
the ratio $c_{4}/\chi_{\rm t}$ converges to a value close to $-1$ 
\cite{Lat16} (which is the value of a dilute instanton gas).

\item If we add a $\theta$-term,
  $S[\vec e \, ]_{\theta} = S[\vec e \, ] - \ri \theta Q[\vec e \, ]$,
  with $-\pi < \theta \leq \pi$,
  we obtain an expectation value $\la Q \ra$, which does not need to
  vanish anymore. Therefore we now have to refer to the general
  expression for $\chi_{\rm t}$,
\bea
\la Q \ra &=& - \ri \partial_{\theta} \ln Z (\theta) \ , \nn \\
\chi_{\rm t} &=& \frac{1}{V} \Big( \la Q^{2} \ra - \la Q \ra^{2} \Big)
  = -  \frac{1}{V} \partial_{\theta}^{2} \ln Z (\theta) \ .
\eea
  The expectation value $\la Q \ra$ is well-defined at
  any vacuum angle $\theta$, but the
  function $\la Q \ra (\theta)$ has an infinite slope
  at $\theta = 0$. This is the picture elaborated in
  Ref.\ \cite{Bogli}, without GF, which is sketched schematically
  in Figure \ref{Qtheta}. It implies that $\theta$ remains finite
  under renormalisation,\footnote{Ref.\ \cite{Bogli} concludes
  that each value of $\theta \in [0, \pi ]$ represents a different
  continuum theory.}
  and that $\chi_{\rm t}(\theta)$ does exhibit continuum scaling 
  at any $\theta \neq 0$.
    
\end{itemize}

We have seen that the picture of Ref.\ \cite{Bogli} --- in
particular the infinite slope at $\theta = 0$ --- seems to
(qualitatively) persist under the GF. This extends our previous
observation \cite{procs} to much longer flow times.

Ref.\ \cite{Bogli} did not consider this behaviour unnatural,
although $\chi_{\rm t} (\theta =0)$ is supposed to be an observable.
This scenario requires the free energy
$F(\theta ) = -\beta^{-1} \ln Z(\theta )$
to take an unusual --- though conceivable --- form,
where $F(\theta )$ and $F'(\theta )$ have removable singularities
at $\theta = 0$, which give rise to a divergence
of $F''(0)$ (a prototype for such a function is
$F(\theta ) \propto \theta^{2} \ln \theta$).

Here we present our numerical results, which support this
scenario. We leave it to the reader to decide whether he/she
considers this property as fatal for the topology of the
2d O(3) model. \\

\ \\
\noindent
{\bf Acknowledgements:} We are indebted to Martin L\"{u}scher for
suggesting this project, and for helpful advice on its realisation.
We further thank Uwe-Jens Wiese for instructive
discussions. The computations were performed on the cluster of
ICN/UNAM; we thank Luciano D\'{\i}az and Eduardo Murrieta for
technical support. This work was supported by DGAPA-UNAM
through grant IN107915 and through the program PASPA-DGAPA,
by the Albert Einstein Center for Theoretical Physics and by
the European  Research  Council under the  
European  Union's  Seventh  Framework Programme 
(FP7/2007-2013)/ERC grant agreement 339220.
PdF thanks the CERN Theory Division for its hospitality.

\appendix

\section{Numerical integration of the Gradient Flow}

This appendix compares various
implementations of the GF based on the Runge-Kutta method; for a 
pedagogical description of this method we recommend Ref.\ \cite{NumRec}.
In particular we are going to address the performance
of the Dormand-Prince adaptive step size algorithm \cite{DoPri}.

That algorithm allowed us to handle lattices up to size $L = 404$ 
with a high statistics of $2 \cdot 10^{5}$ configurations, 
see Tables \ref{tabLbxi} and \ref{tabchit}.
In the smaller volumes we could run the GF at a fixed step size
of $dt = 10^{-4}$, and extensive tests demonstrated the consistency 
with the Dormand-Prince algorithm, up to $L=270$. This appendix 
is going to concentrate on $L=404$, where fixed $dt = 10^{-4}$ 
production runs are prohibitively expensive. Instead
we refer to a sample of 100 test configurations, which were
generated at $\beta = 1.807$ (the value used in our study),
well thermalised and independent.

Our tests have further shown that the most delicate
part of the GF is the very beginning. This is expected:
possible artifacts due to the finite step size $dt$ are most
likely before the configurations become smooth.
In this appendix we consider flow time $t=0$ to 
$10 \simeq 0.35 t_{0}$. This interval is of primary interest:
we will see that most of the reduction of $\la Q^{2} \ra$ that we 
observe up to $10t_{0}$ (see Figure \ref{Q2vsGF}) 
happens in the very first flow period.

Strictly speaking, the application of the Dormand-Prince algorithm
requires two parameters: the initial time step $dt_{0}$, and a
``tolerance parameter'' $\vep$. If the gradients computed
by the Runge-Kutta method with 4 points and with 5
points\footnote{Referring specifically to these two gradients is
motivated by the fact that some ingredients of their computation
are identical.}
coincide within this tolerance, {\it i.e.}\ the norm of their
difference is below $\vep$, then $dt$ will be increased in the
subsequent step --- in the opposite case it will be decreased.\footnote{This
is untypical, since the configurations become gradually smoother under
the GF, but it does occasionally happen, {\it i.e.}\ the increase of
$dt$ is not strictly monotonous.\label{fnstepsizedown}}

Regarding the initial time step $dt_{0}$, we ran numerous tests
with $dt_{0}=10^{-3}$ and $dt_{0}=10^{-4}$: when everything else was
kept fixed, we never found any difference which could be significant
at our level of precision. After just a few time
steps one obtains results, which are practically indistinguishable.
Since this choice hardly affects the computation time,
we used $dt_{0}=10^{-4}$ in our production runs,
and also in the tests to be presented in this appendix. Hence our
discussion focuses on the tolerance parameter $\vep$.

We are going to compare three numerical implementations of the GF:
\begin{itemize}
\item Fixed step size $dt=10^{-4}$.
\item Dormand-Prince adaptive step size with $dt_{0}=10^{-4}$
and $\vep = 10^{-6}$ (as used in our production runs).
\item The same Dormand-Prince algorithm with $\vep = 10^{-7}$. 
\end{itemize}

\begin{figure}[h!]
\begin{center}
\includegraphics[width=0.45\textwidth,angle=270]{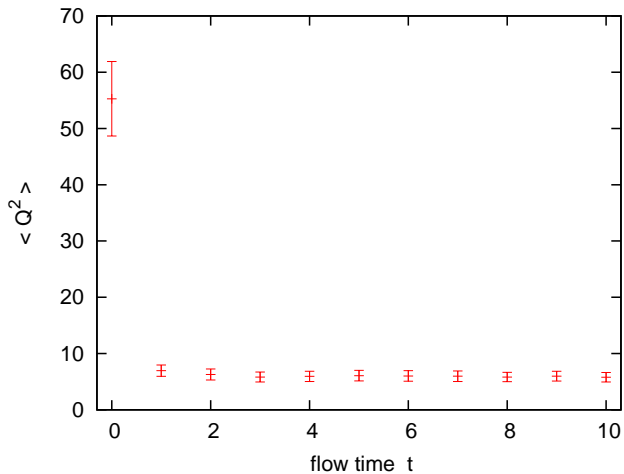}
\end{center}
\vspace*{-7mm}
\caption{The mean value $\la Q^{2} \ra (t)$, obtained from 100
configurations at $L=404$, at flow times $t = 0 \dots 10 t$. 
The results coincide for all three GF implementations under 
consideration.}
\label{Q2test}
\end{figure}

First we consider the topological charges of these 100 test 
configurations. We checked for possible deviations when we apply 
these GF implementations, but
they fully agree at any $t = 1,\, 2,\, 3 \dots 10$.
Figure \ref{Q2test} shows the value of the $\la Q^{2}\ra$ obtained
from this sample.
It confirms that most of the destruction of topological windings
happens very early, at $t<1 \simeq 0.035 t_{0}$. 
This corresponds to an impact range below $2$ lattice spacings, 
hence it matches the picture of a quick destruction of numerous 
tiny dislocations (compared to the correlation length $\xi \simeq 68$).
The topological windings that persist can either be large, or small
with a structure, which resists the GF for longer flow time. 
We saw that these remaining windings still make the topological
susceptibility diverge.

\begin{figure}[h!]
\begin{center}
\includegraphics[width=0.45\textwidth,angle=270]{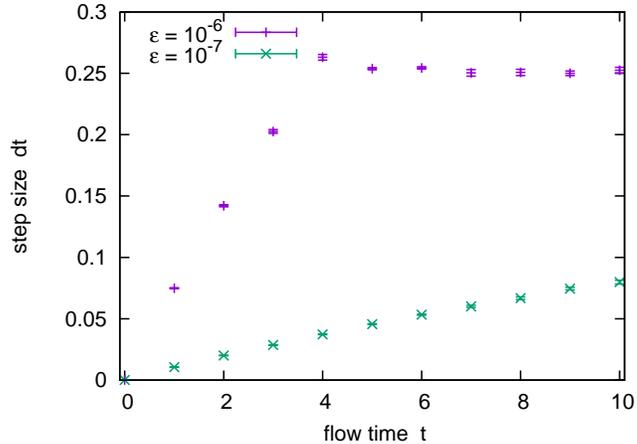}
\end{center}
\vspace*{-7mm}
\caption{The step size evolution of the Dormand-Prince algorithm
with an initial time step $dt_{0} = 10^{-4}$, and with
tolerance parameter $\vep = 10^{-6}$ or $10^{-7}$.}
\label{timestep}
\end{figure}

Figure \ref{timestep} illustrates how $dt$ increases when we apply
the Dormand-Prince algorithm, with tolerance parameter
$\vep =10^{-6}$ or $\vep =10^{-7}$. The difference between these 
two scenarios is significant: in particular, at 
$\vep =10^{-6}$ the step size soon attains a remarkable magnitude
of $dt \approx 0.25$; at $t \approx 4$ the configurations
are already sufficiently smooth to allow for this value. 
(That case also confirms that, in exceptional cases, 
the algorithm can temporarily decrease $dt$, cf.\ 
footnote \ref{fnstepsizedown}.)

Since we did not observe any significant difference in the results, 
the use of this value of $\vep$ is highly motivated. It provides
a gain in computation time by several orders of magnitude: this gain
can be estimated by assuming the GF to take computation time 
$\propto 1/dt$, although adaptive step size algorithms require
some additional operations.

\end{document}